# Revisiting Christoph Scheiner's Sunspot Records: a New Perspective on Solar Activity of the Early Telescopic Era

V.M.S. Carrasco[1,2], A. Muñoz-Jaramillo[3], M.C. Gallego[1,2], J.M. Vaquero[2,4]

[1] Departamento de Física, Universidad de Extremadura, 06006 Badajoz, Spain [e-mail: vmscarrasco@unex.es]

[2] Instituto Universitario de Investigación del Agua, Cambio Climático y Sostenibilidad (IACYS), Universidad de Extremadura, 06006 Badajoz, Spain

[3] Southwest Research Institute, Boulder, CO 80302, USA

[4] Departamento de Física, Universidad de Extremadura, 06800 Mérida, Spain

**Abstract:** Christoph Scheiner was one of the most outstanding astronomers in the history of the sunspot observations. His book, *Rosa Ursina*, is the reference work regarding the study of the earliest sunspot records. The sunspot observations compiled by Scheiner in *Rosa Ursina* and *Prodomus*, including records made by other observers, forms one of the main references of the observations known for that period; particularly around the 1620s. Thus, his work is crucial to determine the solar activity level of the first solar cycles of the telescopic era. The number of sunspot groups recorded in Scheiner's documentary sources has been included in the existing sunspot group number databases. However, we have detected significant errors in the number of groups currently assigned to Scheiner's records. In this work, we reanalyze the information in Scheiner's source documents. Consequently, the standard 11-yr solar cycle shape for the second solar cycle of the telescopic era, which is not clear in previous studies, now becomes evident. In addition, the highest daily number of groups recorded during this cycle (8 groups) is 20% less than the one included in the existing sunspot group number databases. Using the hypergeometrical probability distribution, we find that solar minima in 2008-2009 and 2018-2019 are comparable to the most probable solar activity level of the minimum around 1632. In particular, the estimated lower limit for the solar activity in 1632 is even comparable with the solar activity level in 2008 and 2018.

## 1. Introduction

The first known sunspot observations were made by ancient civilizations using the naked eye (Hardy 1991; Keimatsu 1970, Vaquero & Vázquez 2009). Approximately, a few hundreds of naked-eye sunspot records made during the last two millennia are available (Wittmann & Du 1987, Yau & Stephenson 1988, Xu et al. 2000, Willis & Stephenson 2001, Vaquero et al. 2002, Hayakawa et al. 2017, Willis et al. 2018). Sunspot observations were recorded more systematically when telescopes started to be used as an astronomical instrument since 1610 (Muñoz-Jaramillo & Vaquero 2019, Arlt & Vaquero 2020, Vokhmyanin et al. 2020). A complete daily observational coverage of sunspot records is roughly available since mid-19th century (Hoyt & Schatten 1998, Clette et al. 2014, Vaquero et al. 2016). Nowadays, more than 400 years later, sunspot counts constitute the longest direct observation set of solar activity. Moreover, it is considered as the longest-lived running "experiment" in the world (Owens 2013).

Galileo and Scheiner disputed who was the first to discover sunspots (Galileo & Scheiner 2010). However, the first well-dated sunspot observations by telescope available in the current sunspot group number database (Vaquero et al. 2016, hereafter V16) were recorded by Harriot in December of 1610 and Fabricius in March 1611 (Herr 1978, Chapman 1995, Neuhäuser & Neuhäuser 2016, Vokhmyanin et al. 2020). Other astronomers who recorded sunspot observations in that time were Malapert, Jungius, and Mögling. Several works have been recently published using these early observations shed light on solar activity during the first years of the telescopic era as, for example, Carrasco et al. (2020) who identified a significant number of sunspot observations made by Galileo and Scheiner not included in V16. Moreover, these authors showed how the observation method by projection invented by Castelli (a student of Galileo) and used by Galileo from 3 May 1612 had an important impact in the number of groups recorded by this astronomer. Arlt et al. (2016, hereafter A16) and Vokhmyanin & Zolotova (2018a) derived the positions and group number of sunspots recorded by Scheiner (1630) and Galileo in their drawings, respectively. Neuhäuser & Neuhäuser (2016) revised some early sunspot observations made by several astronomers, such as Marius and Saxonius, and Carrasco et al. (2019a) pointed out some problematic number of groups included in V16 due to misinterpretations of Malapert's sunspot observations. Moreover, Carrasco et al. (2019b) found the standard shape of the 11-yr solar cycle in the solar cycle around 1620s using Malapert's data in addition to a reduction of the number of groups assigned to this observer in V16. Hayakawa et al. (2021) analyzed Mögling's original manuscripts to revise his sunspot group number and derive sunspot positions, revising Schickard and Hortensius' sunspot group number. We also note the work carried out by Vokhmyanin & Zolotova (2018) and Carrasco et al. (2021a) on the group number, sunspot positions and areas recorded by Gassendi in the 1630s. Carrasco et al. (2019c) analyzed the sunspot observations made by Hevelius just before the Maunder Minimum and obtained that the solar activity level computed from those Hevelius' sunspot observations is significantly greater than that according to the sunspot records made by Hevelius during the Maunder Minimum. In addition, Hevelius recorded a period of around three months without sunspots at the end of 1644 and early 1645 (Carrasco et al. 2019c) providing an indication of the possible start of the Maunder Minimum (Eddy 1976; Usoskin et al. 2015).

However, despite the latest efforts to improve our knowledge of solar activity in the first part of the telescopic era, further improvements are still required: the observational coverage for the two first solar cycles of the telescopic era is currently around 20% of the days. It is also important to study these early sunspot observations in order to know how the transition of solar activity changed from a normal regime to a grand minimum period (Vaquero et al. 2011). Considering these objectives, we revise the sunspot observations published by Scheiner in *Rosa Ursina* (Scheiner 1630) and *Prodomus* (Scheiner 1651) after detecting some misinterpretations and missing information in previous works on these sources (Hoyt & Schatten 1998, Vaquero et al. 2016, Arlt et al. 2016). We provided some notes about Scheiner and their previously mentioned documentary sources in Section 2. We analyze and discuss the new sunspot group counting made in this work from Scheiner's sunspot observations in Section 3. In Section 4, we show a comparison between modern minima and the solar activity level of the solar minimum around 1632 computed from the active day fraction. Finally, the main conclusions are exposed in Section 5.

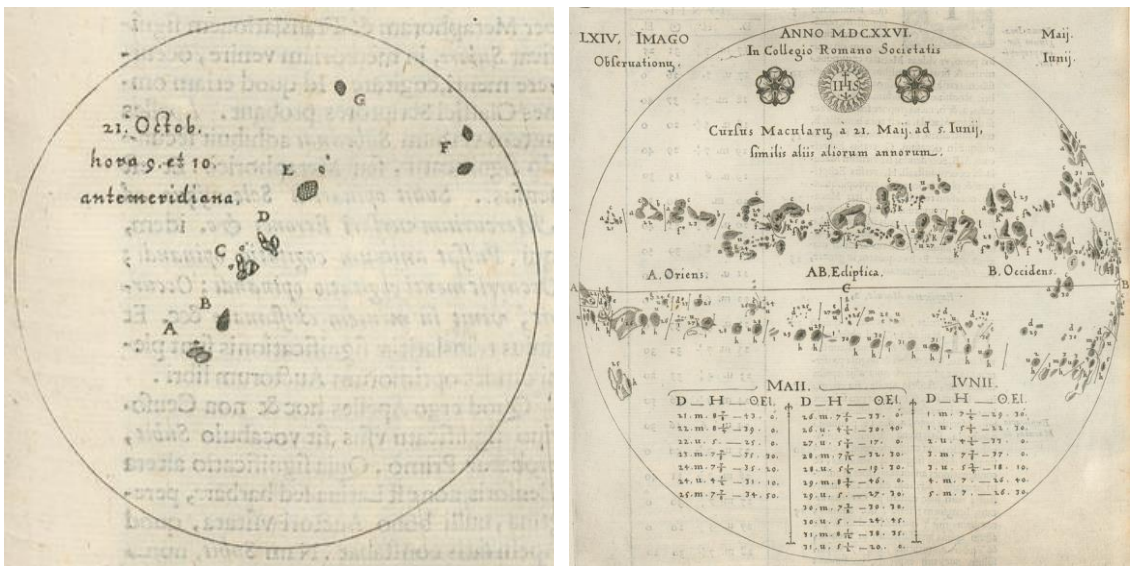

**Figure 1.** Two example drawings with sunspot records made by Christoph Scheiner observing through a telescope with colored glass on 21 October 1611 (left panel) and using the new methodology by projection with improved telescopes from 21 May to 5 June 1626 (right panel) [Source: Scheiner 1630, p. 63 and 315].

**2. Scheiner, *Rosa Ursina*, and *Prodomus***

Christoph Scheiner was born in Walda (Germany) in 1573. He joined the Jesuit order at Lansberg in 1595 and studied Mathematics at Ingolstadt (Mitchell 1916, Daxecker 2004). Scheiner was sent to teach this subject at Dillingen. There he invented the pantograph: a device to reproduce drawings to different scales. In 1610, he returned to Ingolstadt and stayed there until 1616. In this period, he carried out his first sunspot observations. Later, he moved to Innsbruck in 1616 and became a professor of Mathematics at Freiburg in 1620, where he published a study on the optics and physiological nature of the human eye (Scheiner 1621). In 1623, Scheiner was appointed Rector of the Jesuit College at Niesse (today Nysa, Poland) and, later, he was sent to Rome from 1624 to 1633. In this period, he published his most important book in terms of scientific relevance, *Rosa Ursina* (Scheiner 1630), including his sunspot observations made from 1611 to 1627 and carried out the sunspot observations that were published in his posthumous work *Prodomus* (Scheiner 1651). Afterward, he lived at Vienna in 1633 returning to Niesse definitely in 1637, where he passed away in 1650. In addition to the works previously mentioned, he also performed studies in mathematics and sundials.

Because of the Aristotelian ideas of that time, Scheiner firstly attempted to explain his sunspot observations assuming the perfection of the Sun. Thus, he postulated that sunspots were celestial bodies orbiting the Sun. He sent three letters under the pseudonym of Apelles explaining his work to Marcus Welser who published them with the title *Tres epistolae de maculis solaribus*. Galileo refuted these arguments, leading to one of the famous debates in the history of astronomy between Galileo and Scheiner (Shea 1970). Although Scheiner was initially wrong on the nature of sunspots, he was nevertheless a great scientist and a superb observer. His sunspot observations were rigorous and of good quality, in particular those made in the 1620s. Scheiner continued observing sunspots for approximately two decades. Scheiner designed and built different telescopes, including one with convex lenses that improved the solar image in his observations of the 1620s (Arlt et al. 2016). Moreover, he also modified the observation method by projection, used for the first time by Castelli and Galileo in 1612 (Galileo & Scheiner 2010, Carrasco et al. 2020), by adding to the helioscope the earliest known equatorial mount. Scheiner named this instrument the "heliotropic telescope". These two innovations significantly improved the quality of his sunspot observations. Figure 1 depicts two examples of sunspot drawings recorded by Scheiner before and after using the new methodology with improved instruments (Scheiner 1630). Note that the first sunspot observations made by Scheiner were made through the telescope using a colored glass.

Scheiner's two documentary sources analyzed in this work are *Rosa Ursina* (Scheiner 1630) and *Prodomus* (1651). *Rosa Ursina* is divided in four parts. The first one discusses the question of the discovery of the sunspots and also includes Scheiner's admission that his previous conclusions on the nature of sunspots were wrong, in addition to his sunspot observations made in 1611. The second part includes descriptions of the telescopes and the observational method he used. The third part collects sunspot observations made from 1618 to 1627. The fourth and last part deals with sunspots, the solar rotation, and the inclination of its rotational axis. Scheiner used his sunspot observations to demonstrate the inclination by 7º of the solar equator with respect to the ecliptic.

The second book, *Prodomus*, was posthumously published in 1651. This book collects several years of Scheiner's work and his effort to refute the heliocentric theory. It also includes a criticism to Galileo because of his work published in 1632 "Dialogo" (Drake 1967). *Prodomus* contains several drawings published by Scheiner (1630), in addition to new sunspot observations made until the end of 1632.

**3. Analysis and Discussion of Scheiner's Sunspot Observations**

The sunspot observations published by Scheiner (1630, 1651) were previously analysed by Wolf (1859) for the creation of the sunspot number index. Hoyt & Schatten (1998) used observations from these two books to extend the group sunspot number to the beginning of the 17th, among other records, pointing out that Elizabeth Nesme-Ribes also re-examined *Rosa Ursina* to get group counts. V16 incorporated data from Hoyt & Schatten (1998) to a consolidated sunspot group number database. Furthermore, A16 used observations from these books to classify sunspot groups and calculate sunspot positions, areas, and group tilt angles for 1611−1631.

These sources contain observations made in 1611 and also for the period 1618–1632. Most of *Rosa Ursina*'s and *Prodomus*' observations are contained in sunspot drawings although other observations form part of the textual reports. Figure 2 (top panel) shows the number of sunspot groups included in V16 for the period 1610–1649. Red color shows the 882 daily sunspot records attributed to Scheiner in V16 during the periods 1611–1640. However, we have detected several mistakes in the way sunspots recorded by Scheiner in *Rosa Ursina* and *Prodomus* were counted because of significant misinterpretations of the observations. Some of these have been already reported by Carrasco (2019) such as those observations made on 27 February 1622 and 16 January 1625.

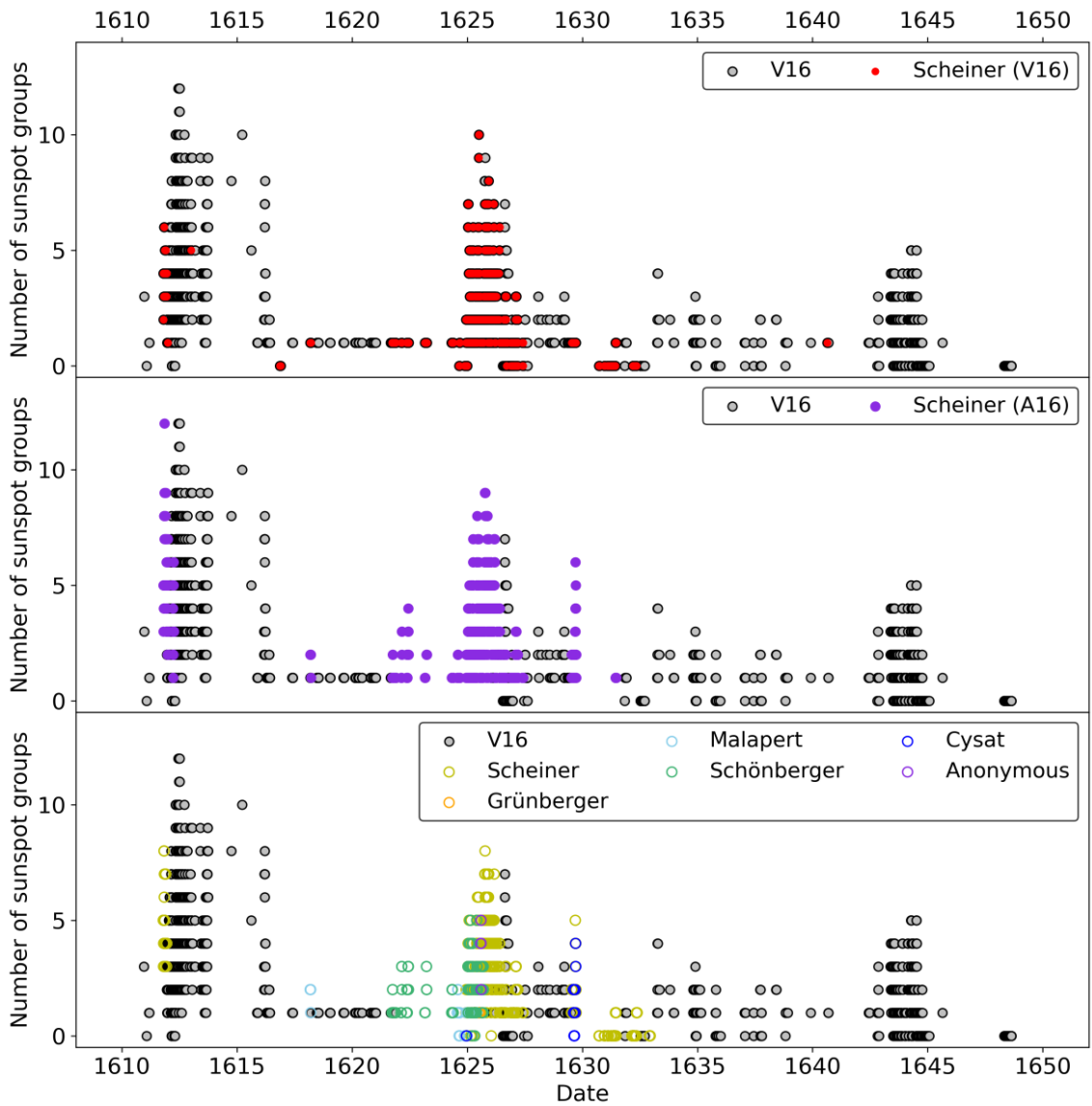

**Figure 2.** Daily number of sunspot groups recorded by all the observers available in the current sunspot group number database by V16 during the period 1610–1649 with Scheiner's observations represented by red color (top panel). Daily number of sunspot groups included in V16 for the period 1610–1649 (grey full circles) removing Scheiner's and Smogulecz's data, in addition to the counts made by Arlt et al. (2016) in purple (middle panel) and the new counts made in this work: Scheiner in yellow color,

Grünberger in orange, Malapert in light blue, Schönberger in green, Cysat in blue, and Anonymous in violet (bottom panel).

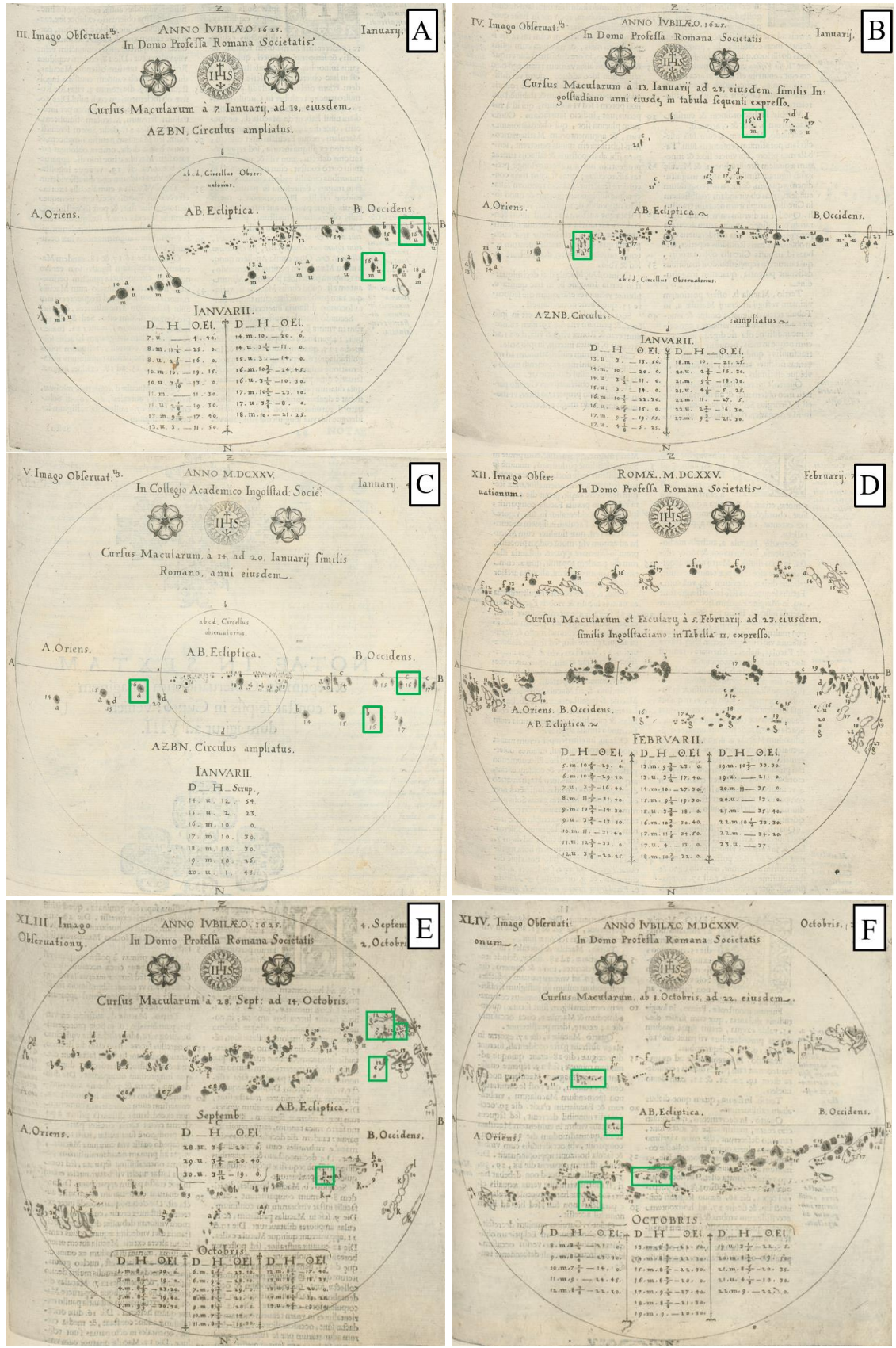

**Figure 3.** Sunspot drawings made by Scheiner (panel A and B) (Scheiner 1630, p. 169 and 171) and Schönberger (panel C) (Scheiner 1630, p. 173) including observations for the period 14 – 17 January 1625. Observations made by the two astronomers on 16 January were marked by green squares. Panel D depicts one of the sunspot drawings made by Scheiner in February 1625 (Scheiner 1630, p. 185) where it can be seen that the big group defined by Scheiner as "b" and "c" was omitted in the drawing on 13, 16 and 18 February although Scheiner observed in those dates. Panel E and F include the sunspot groups (green squares) recorded by Scheiner on 12 October 1625.

### 3.1. Errors Detected in Previous Works

The first kind of error is that the number of groups recorded by others observers is assigned to Scheiner. Scheiner's record (1630, 1651) includes the observations made by six other different astronomers. In Section *Totius operis notatu dignoria*, Scheiner (1630) specifically points out that he also shows drawings from observations made by Grünberger at Rome (Italy), Malapert at Douai (nowadays France), and Schönberger at Freiburg and Ingolstadt (Germany). Scheiner (1630, p. 257) mentions that Georg Schönberger, Scheiner's student and successor in the University of Freiburg, attributes the observations made at Ingolstadt from 4 to 11 August 1625 to another observer (whose name is not mentioned). Scheiner (1651) also included observations made in Vienna (Austria) by Johann Baptist Cysat – a Jesuit astronomer born in Luzern (Switzerland). Neither V16 nor A16 provided the observers responsible of each observation whereas we have indicated that the sunspot observations published in those sources were made by multiple astronomers (Figure 2). An example of this problem occurs on the period of 14–17 January 1625. According to V16, Scheiner reported six groups on the 14 and 15 January and seven groups on the 16–17 January 1625. However, Scheiner only recorded three groups each day for 14–15 January and four groups both on 16 and 17 January (Scheiner 1630, p. 169 and 171). On the same period, Schönberger recorded three groups each day for the period 14–17 January 1625 (Scheiner 1630, p. 173). Counting both observers results in the reported number of groups (mistakenly attributed to Scheiner): six groups for 14-15 January and seven for 16-17 January. These observations are included in Figure 3. As an example, groups observed by Scheiner and Schönberger on 16 January 1625 are pointed out by green squares.

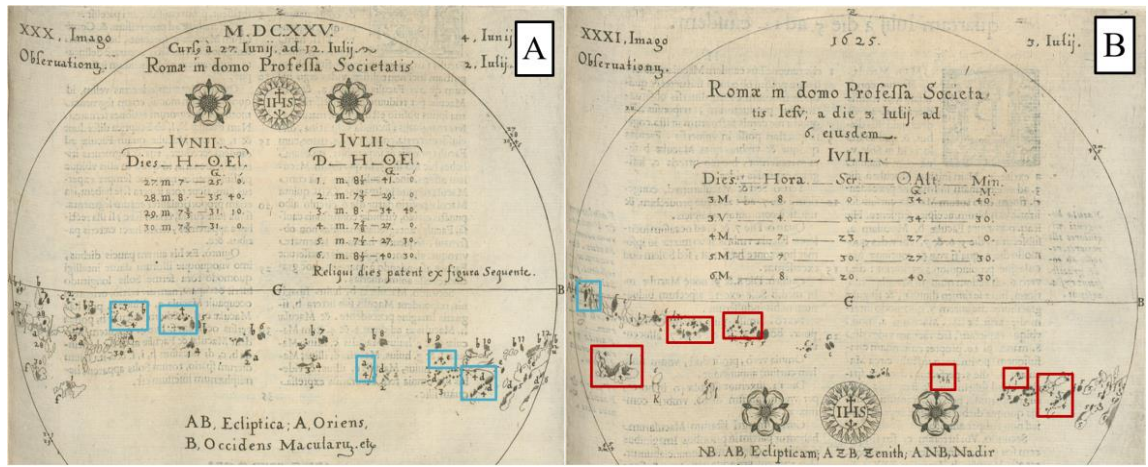

**Figure 4.** Sunspot drawings (A) XXX (Scheiner 1630, p. 235) and (B) XXXI (Scheiner 1630, p. 237) recorded by Scheiner with information on the sunspot groups recorded on 4 July 1625. Blue and red squares depict the solar phenomena recorded in that date. Red squares represent the same groups recorded by blue squares except the solar facula recorded by Scheiner with the letter "l".

The second kind of error found in V16 relates to double counting of groups drawn twice in a single day. We note that the group counting by Arlt et al. (2016) is free of this error. One striking example of this kind of error is the highest number of sunspot groups included in V16 in the 1620s. According to V16, Scheiner recorded ten groups on 4 July 1625. However, in reality these numbers reflect the double counting groups that were drawn in two separate drawings made on the same day (Figure 4): drawing XXX (Scheiner 1630, p. 235) and drawing XXXI (Scheiner 1630, p. 237). Frequently, Scheiner recorded groups observed on the same dates in different drawings because of the lack of space to include new groups in the drawing. All the solar phenomena (sunspots and faculae) recorded by Scheiner on 4 July 1625 are identified in Figure 4 by squares. There are five sunspot groups recorded by Scheiner in Figure 4A and six groups in Figure 4B. Note that the solar phenomenon "l" furthest to the left in the red squares (Figure 4B) is a solar facula, not a sunspot group. Nevertheless, five groups identified by red squares in Figure 4B are the same groups as those recorded in Figure 4A. Thus, it is incorrect to include the five groups in red squares to determine the group counting because they are the same groups already counted in Figure 4A. This means that the real number of groups recorded by Scheiner on 4 July 1625 is not the sum of all of groups recorded in Figure 4. Only the six groups identified by blue squares in Figure 4 count.

Finally, we noticed that Scheiner may have omitted some groups in his drawings since these groups were observed in previous and subsequent days to the date when they were not recorded. For example, a great group composed by sunspots "b" and "c" according to Scheiner (1630, p. 185) from 9 to 22 February 1625 was not recorded on 13, 16, and 18 February although Scheiner made observations in these dates (Figure 3D). Given its size, this large group should have been visible to Scheiner on the 13, 16, and 18 February (as in previous and subsequent days). We found this kind of possible error in other 14 observation days. We hypothesize that this is likely caused by lack of drawing space. Carrasco et al. (2019) found a report of this problem in the Latin translation of Scheiner (1630, p. 195) according to Malapert's observations from 24 March to 2 April 1625. There, Scheiner indicated that not all sunspots were recorded in the drawings due to the space limitation in the paper. We can find other example comparing the sunspot observations recorded by Scheiner for the period 1–7 February 1625, divided into two drawings (Scheiner 1630, p. 177 and 183), with those made by Schönberger (Scheiner 1630, p. 179) and Malapert (Scheiner 1630, p. 181) for the same dates. The great sunspot "a" recorded by Scheiner (1630, p. 183) from 1 to 13 February 1625 is not included in the drawings by Schönberger and Malapert. Because of the significant size of the sunspot "a" is very probable that both astronomers observed this spot. This could suggest that Scheiner had a particular interest in comparing the trajectories of the groups recorded only in the drawing VIII (Scheiner 1630, p. 177) with those equivalent groups observed by Malapert and Schönberger. In fact, this group is included in other drawing by Schönberger (Scheiner 1630, p. 199) but never by Malapert.

### 3.2. A Recount of Group Numbers

We recounted the number of sunspot groups in 1083 observations for the period 1611–1632 from Scheiner (1630, 1651). These records correspond to 928 different observation days. Scheiner is the observer with the highest number of daily records (814) regarding the observations included in those documentary sources followed by Schönberger (166), Malapert (62), Cysat (22), Grünberger (13), and the anonymous astronomer cited above (6). We recounted the number of groups recorded in their sunspot observations according to the modern classification of sunspot groups (McIntosh 1990). We highlight that no additional group was added to the group counting in those days when an evident omission of some group was detected, except in the case showed by Carrasco et al. (2019) according to Malapert's observations for 24 March – 2 April 1625.

Figure 2 (bottom panel) shows our new counting of sunspot groups carried out in this work from Scheiner (1630, 1651). Our counting is publicly available at the website of the Historical Archive of Sunspot Observations (http://haso.unex.es). Note that: i) the sunspot records made by Smogulecz for the period 1621–1625 were omitted in Figure 2 (middle and bottom panel) because the number of groups included in V16 according to this observer are wrong due to misinterpretations of the observations (Carrasco et al. 2021b), ii) we incorporate 108 additional observation days not included in V16 according to Scheiner (1630, 1651), and iii) we also find that 44 observation days in V16 from Scheiner (1630, 1651) are wrong because no observation was made in those dates.

The highest daily number of sunspot groups recorded by any observer for the second solar cycle of the telescopic era was 10, according to V16. This record would have been made by Scheiner on 4 July 1625. However, we have shown that it is wrong because of the misinterpretation of the groups recorded in the drawings (Figure 4). The highest number of groups recorded in that period, according to this new counting, is that made by Scheiner on 12 October 1625 when he recorded 8 groups (Figure 3, panel E and F). These groups are recorded in drawings XLIII and XLIV in Scheiner (1630, p. 265 and 267). This implies a decrease of 20% in the maximum amplitude for that solar cycle with respect to the previous knowledge.

Between 1616 and 1624, no sunspot records have a number of groups greater than one according to V16. However, the number of groups from this new counting is up to 3 in that period. Now, the ascending phase of that solar cycle has a more standard shape and not an abrupt change from the minimum to maximum (see for example Figure 30 in Svaalgard and Schatten 2016). A similar behavior happens in the declining phase. We have counted up to 5 groups recorded by Scheiner in 1629 (September 12) when it was only 1 in V16. This results on a more gradual decrease of solar activity in the declining phase. Thus, the minima of this solar cycle would be around 1620, the maximum amplitude around 1625 and the following minimum around 1632. This result agrees with the standard solar cycle shape found by Carrasco et al. (2019b) and Hayakawa et al. (2021) for the same solar cycle from a revision of Malapert's and Mögling's data.

The solar activity level resulting from our recount of Scheiner's observations at the end of 1611 is significantly greater than that from V16. The average number of groups calculated from V16's Scheiner's observations for the period 21 October – 14 December 1611 is equal to 3.8, while in our recount it is 5.0; i.e. around a 33% higher. Regarding the sunspot observations published by Scheiner (1630, 1651) from 1618 to 1629, the average of the number of groups computed from this work is 2.3 which is very similar to that from V16 (2.2); even though the maximum amplitude of this solar cycle is 20% lower in our recount.

We also compared our group counts with the number of groups defined by A16 from Scheiner's observations published in *Rosa Ursina* and *Prodomus* (Figure 2). The group counts made in both studies are similar in each observation day. However, some differences can be found due to the way to count some sunspot groups (especially in the first sunspot observations by Scheiner). The daily average of the group number corresponding to the sunspot observations published by Scheiner for the period 21 October – 14 December 1611 computed from A16 is 5.8 and it is 5.1 from our group counting using the same observations days as A16. Moreover, the daily group average according to this work and A16, regarding the same observation days for the period 1611–1631, is 2.5 and 2.7, respectively. Note that in the days in which observations from different astronomers are available, we only chosen Scheiner's sunspot records to perform the previous calculations. The highest number of groups counted by A16 was: i) 12 groups in the first solar cycle of the telescopic era on 7 November 1625, and ii) 9 groups in the second solar cycle for 8 – 12 October 1625. We counted up to 8 group in both solar cycles: 1–2 November 1611 for the first solar cycle and 12 October 1625 for the second solar cycle. Occasionally, it is not straightforward to define groups in the first sunspot drawings made by Scheiner, such as that drawing corresponding to 7 November 1611. The way to define several groups included in that sunspot drawing is the main reason of the significant difference in the highest number of groups counted by A16 and this work. Regarding the highest number of groups in the second solar cycle (12 October 1625), we consider that sunspots named by Scheiner as "b", "f", and "g" are the same group (see Figure 3, panel E), while A16 counted it as different groups. Moreover, we provide 332 observations in 177 observation days not included in A16. Note that we recover the spotless days recorded by Scheiner (1630, 1651) whereas A16 did not consider them. There are a few isolated significant differences according to the group counts made in A16 and this work. A difference between group counting made in both studies larger than 3 is only found in five observation days: 7 and 8 November 1611, 28 May 1625, 8 June 1625, 7 March 1626.

In addition to the observations studied in this work, Scheiner published more sunspot records. For example, his sunspot observations made from mid-December 1611 to the beginning of 1613 were studied by Carrasco et al. (2020). Moreover, V16 assigned spotless days to Scheiner on 13–15 and 22–23 November 1616. The description of these records can be consulted in Scheiner (1617, p. 63 and 65–66). After translating the original Latin text, we conclude that Scheiner is not describing sunspots in these observations. One example of this fact can be found in the annotation made by Scheiner on 22 November 1616: (English translation) "The Sun moderately came out calm. It was seen that spots [maculae] had appeared and then the Sun rose over the horizon. When the Sun touched the horizon, the entire periphery could be seen not very torn, and the Sun was cut in half by a stretch of clouds". Thus, Scheiner mentions the term "spots" (maculae) to describe likely the meteorological conditions or the state of the atmosphere just before seeing the Sun over the horizon. That discards that Scheiner recorded a sunspot observation on that date. Thus, these observations should be removed in future versions of the sunspot group number databases. In addition, V16 include from Hoyt & Schatten (1998) two sunspot observations made by Scheiner on 21 and 22 August 1640. We did not find the information corresponding to those observations made in 1640. Thus, more research on these records should be done in the future.

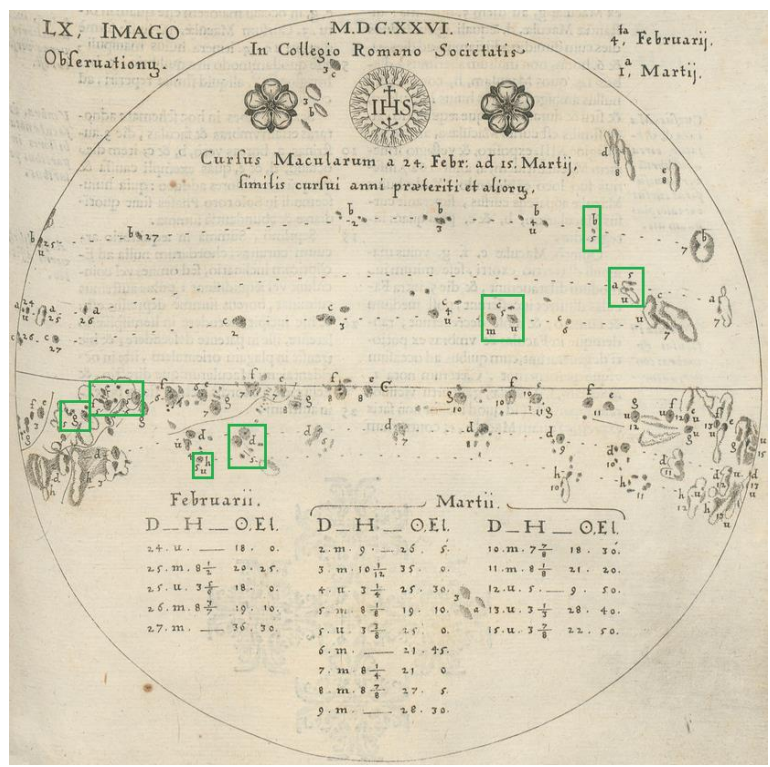

**Figure 5.** Sunspot drawings LX (Scheiner 1630, p. 303) recorded by Scheiner with information on sunspot groups recorded on 5 March 1626. Green squares represent the seven groups defined in this work for this date.

**3.3. Scheiner, Schönberger, and Malapert: a Comparison**

Some of the sunspot records made by other observers, published by Scheiner (1630, 1651), were carried out in dates that Scheiner also observed. Thus, we can perform a direct comparison between observations of several astronomers. Figure 6 represents the daily number of groups recorded by Schönberger (top panel) and Malapert (bottom panel) against those made by Scheiner in the same dates. Schönberger and Scheiner made observations in 97 common dates from 14 January to 15 September 1625. In all those days, Scheiner recorded more groups than Schönberger in 41 observations days and Schönberger only in 9 days more than Scheiner. The average of the number of groups recorded by Schönberger in those days is 2.1 while it is 2.6 in the case of Scheiner. Malapert and Scheiner recorded observations in 18 common dates. The average of the number of groups computed from Malapert's observations in those days is 2.3. In the case of Scheiner, the average is equal to 2.8. Given that Malapert only recorded more groups than Scheiner in one of those 18 days, Scheiner systematically recorded more groups than Schönberger and Malapert.

We also carried out a comparison between observations by other observers with a lower number of common observations days. For example, Scheiner and Cysat recorded observations in 15 common dates during the periods 11–22 August and 10–19 September 1629, including one spotless day (21 December 1624) as well. The average of the number of groups obtained from Scheiner's records is 1.9, similar to that from Cysat's data for those date which is 1.7. Furthermore, Schönberger and Malapert recorded a similar number of groups in their common observation days. They observed in 12 common dates. The average of the number of groups calculated from Schönberger in those days is 2.3 and it is 2.1 according to Malapert's data.

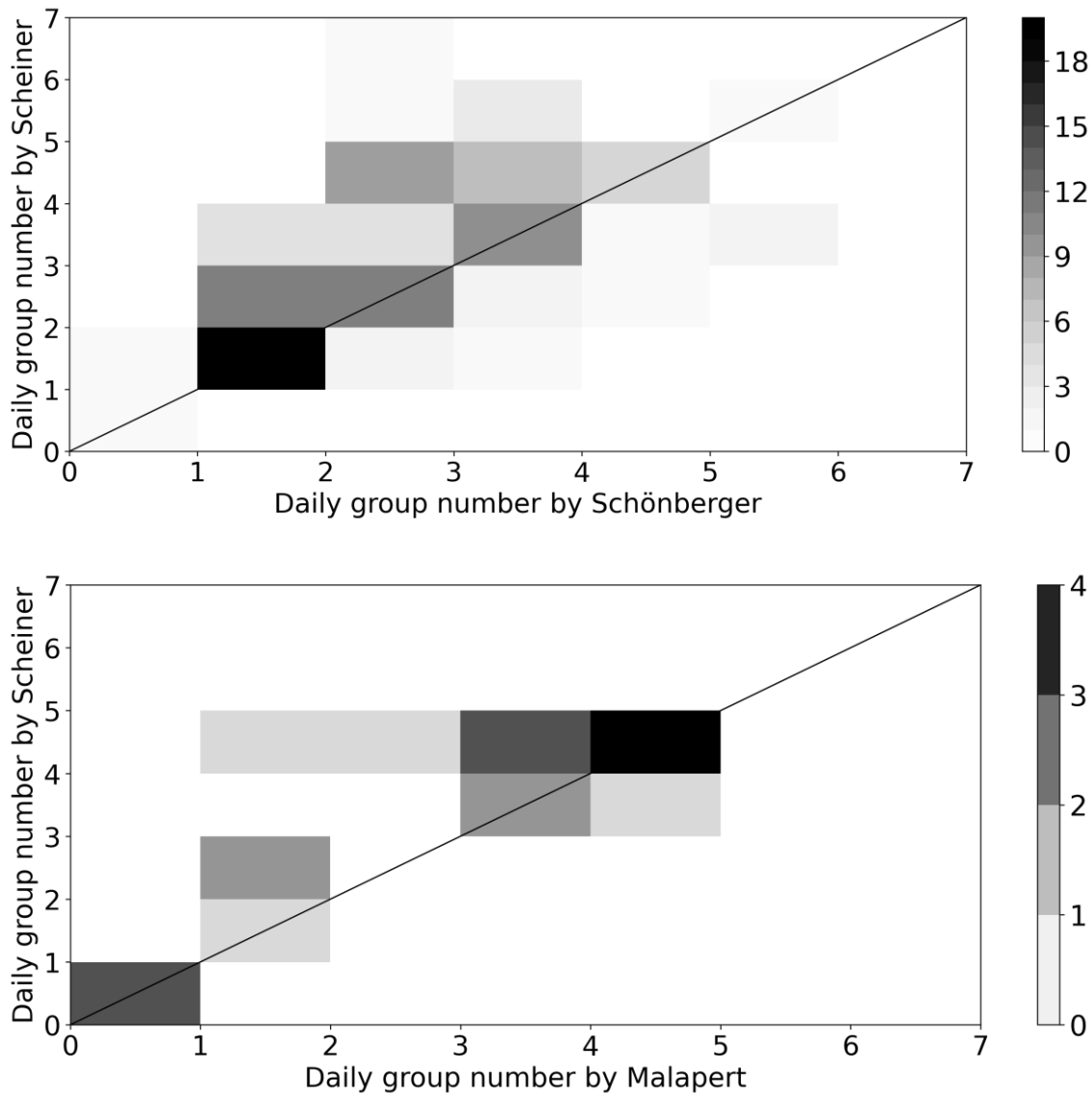

**Figure 6.** Daily number of sunspot groups recorded by Schönberger (top panel) and Malapert (bottom panel) versus Scheiner. Different colors represent the frequency occurred in each combination. Note that only common observation days were taken into account. The diagonal line represents the slope line equal to one.

## 4. Active day fraction around the minimum of the 1630s

Scheiner (1651) recorded a large number of quiet days (days without any spot on the Sun) for the period 1630–1632. He also indicated other significant set of days when he observed sunspots in the same period but without specifying the number of groups or single sunspots. We consider these days as active days (days with at least one sunspot on the visible solar disc). In particular, Scheiner provided information on active and quiet days in the following periods (Scheiner 1651, p. 53-56): i) 12 September 1630 – 22 June

1631, ii) 26 February – 8 March 1632, iii) 13 March – 11 April 1632, iv) 1 – 20 May 1632, and v) 1 – 13 December 1632.

Scheiner observed the Sun almost every day for the period 12 September 1630 – 22 June 1631. In this period, Scheiner (1651) recorded 269 observations days with 65 quiet days and 204 active days. There are only 15 days without observation for that whole period. Thus, the active day fraction in this period is equal to 75.8%. If we suppose that those 15 days were active, then the active day fraction would be of 77.1% and 71.8% in case those 15 days were quiet. This solar activity level is similar to that found from sunspot data provided by the Sunspot Index and Long-term Solar Observations (SILSO, http://www.sidc.be/silso/) for 2006 (82.2%) and 2017 (73.7%), both in the last part of the declining phases of the modern Solar Cycle 23 and 24, respectively. Note that according to sunspot number index provided by SILSO, the minima of Solar Cycle 24 and 25 were in December 2008 and 2019, respectively.

The solar minimum of the third solar cycle regarding the telescopic observations from 1610 was likely in 1632. Several comments made by Scheiner (1651, p. 55–56) points out the prolonged absence of sunspots in that year. The only period of active days recorded by Scheiner in 1632 when he provided specific dates of observation was from 10 to 20 May. Furthermore, according to Scheiner (1651, p. 55), although no sunspot was observed for the most part of January, "some insignificant sunspots, which were not valid for our purpose" were observed in that month. We also note that, from 12 April to the end of the month, "there was hardly any very small sunspot". As a general comment, he pointed out that February was even less active than January and March and April did not bring an increase in activity. In total, Scheiner recorded 63 quiet days and 11 active days for the period 26 February – 13 December 1632. We can use the hypergeometrical probability distribution in order to estimate the solar activity level from the active day fraction for that solar minimum as it follows:

$$p(s) = \frac{s!\,(N-s)!}{(s-r)!\,(N-s-n+r)!} \frac{n!\,(N-n)!}{(n-r)!\,N!\,r!}$$

where $N$ is the number of days in a year (365), $s$ is the total number of active days within the year which is to be estimated, $n$ the number of observations of the sample and $r$ the number of active days of the sample (see more details in Kovaltsov et al. 2004). Thus, we obtained that the most probable value and the upper and lower limits of the active day fraction with a 99% significance level are 14.8%, 23.9%, and 5.7%, respectively. Regarding SILSO's data, only years 2019 (24.9%), 2008 (27.6%), and 2009 (28.2%), which are the solar minima of Solar Cycle 24 and 25, are comparable to the upper limit of the active day fraction obtained for 1632. Figure 7 depicts the active day fraction computed from SILSO data and the most probable value and limits of the active day fraction obtained in this work for 1632.

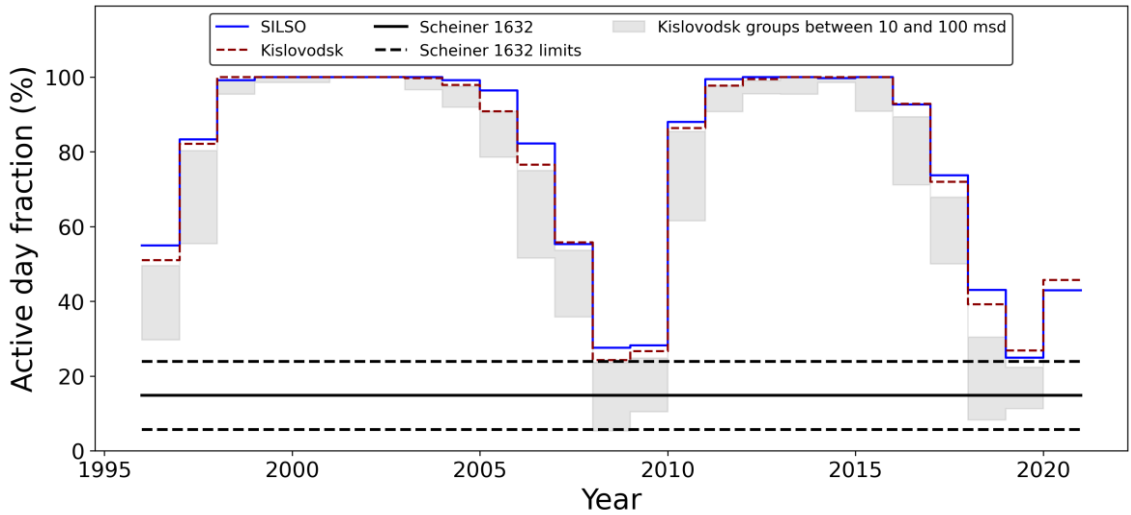

**Figure 7.** Annual active day fraction calculated according to raw data recorded at SILSO (blue line) and Kislovodsk Observatory (red dashed line) for the period 1996–2020. Grey bars represent the annual active day fraction range computed from Kislovodsk data discarding groups whose observed areas are between 10 and 100 millionths of solar disc. Black horizontal lines represent the most probable value of the active day fraction obtained from Scheiner's sunspot records in 1632. Upper and lower limits with a 99 % significant level are depicted by horizontal black dashed lines.

We must take into account that SILSO data are obtained from observations with a modern telescope. Therefore, we should apply some constraint to modern data in order to level with the earliest observations. For example, we can apply area thresholds to modern sunspot records. Thus, we have used the sunspot observation made in Kislovodsk Mountain Astronomical Station of the Central Astronomical Observatory at Pulkovo (Otkidychev & Skorbezh 2014, Muñoz-Jaramillo et al. 2015, Tlatov et al. 2019, Mandal et al. 2020) for the last two solar cycles corresponding to the period 1996–2020 published at: http://158.250.29.123:8000/web/Soln_Dann/. We have calculated the annual active day fraction range from raw Kislovodsk data as well as discarding groups whose observed areas are between 10 and 100 millionths of solar disc (Figure 7). We highlight that the threshold of 10 millionths of solar disc would be an optimistic threshold to be applied to the earliest observations while 100 millionths of solar disc would be conservative (Usoskin et al. 2016). One small difference can be seen between results obtained from SILSO and raw Kislovodsk data. The solar minimum in 2008 is deeper than that in 2019 from Kislovodsk data (2008: 24.3%, 2019: 26.8%), while 2019 is slightly more active than 2008 from SILSO data (2008: 27.6%, 2019: 24.9%). Applying the area thresholds to Kislovodsk data, the minima of Solar Cycle 24 in 2008 (5.3–23.7 %) and Solar Cycle 25 in 2019 (11.3–22.3 %) have a similar solar activity level as 1632. In addition, the solar activity level obtained for 2009 (10.4–24.8 %) and 2018 (8.3–30.4 %) is also comparable to that in 1632. We highlight that the solar activity level in 2008 and 2018 according to the most conservative constraint applied in this work is comparable to the lower limit of the solar activity level estimated for 1632.

## 5. Conclusions

Christoph Scheiner made sunspot observations in the earliest period of the telescopic era. These observations were analyzed previously by Wolf (1859) and Hoyt & Schatten (1998). Then, V16 incorporated the number of groups indicated in these works into the updated sunspot group number database. We have detected several significant errors in the number of groups assigned to Scheiner in V16. For that reason, a revision of sunspot observations recorded by Scheiner was necessary. We have reanalyzed the observations published by Scheiner in his documentary sources *Rosa Ursina* and *Prodomus* from 1611 to 1632. The sunspot records included in these historical works are very important because they are the vast majority of the observations available in the databases, particularly, for the 1620s.

The number of observations studied in this work were 1083 included in more than 100 sunspot drawings made by several observers (Scheiner, Schönberger, Malapert, Cysat, Grünberger, and an anonymous astronomer) in 928 different observation days. We found 108 observation days not included in V16. Moreover, no observation was recorded by Scheiner in 44 observation days assigned to him in V16. Other errors in V16 from Scheiner's data are related to the number of groups recorded by the observers. The most striking case is the highest daily number of groups in the 1620s (second solar cycle known from telescopic observations). According to V16, Scheiner observed 10 groups on 4 July 1625, but in reality he recorded 6 groups. The new highest daily number of groups for that solar cycle obtained in this work was 8 groups recorded by Scheiner on 12 October 1625, i.e. the raw maximum amplitude was 20% lower. In addition, we obtained that the average of the number of groups from Scheiner's observations recorded in 1611 was 5.0 while it is 3.8 using the same dataset from V16, that is around one third greater. Our comparison of observations of different astronomers, also show that Scheiner generally recorded more groups that the other astronomers who recorded observations in the same dates.

We also compared our group counts with the number of groups defined by A16 using the same documentary sources. The group counting is similar in both works but: i) A16 did not provide information on the observers responsible of the sunspot records, ii) no spotless day was included in A16, iii) we provide 177 additional observation days not included in A16 (332 sunspot records if we also consider observers different to Scheiner), and iv) the way to define some particular groups is the reason of the differences found in both works according to the highest number of sunspot groups counted from Scheiner's drawings in the first and second solar cycle of the telescopic era.

We know that Scheiner made a high number of observations around the maximum of second solar cycles (511 observation days for 1625–1626) while he would have carried out his observations just before the maximum of the first solar cycle according to all observations recorded in V16. However, the maximum daily number recorded by Scheiner in each solar cycle was 8 groups: on 1 and 2 November 1611 and 12 October 1625. It was even larger in the first solar cycles according to the group counting by A16 (12 groups in the first cycle versus 9 groups in the second one). Moreover, Scheiner made his observations in the first cycle through the telescope with colored lens and he used the projection method as well as improved telescopes, which could allowe him to observe more sunspot groups (Arlt et al. 2016, Carrasco et al. 2020), during the second cycle. This fact may suggest a gradual decrease of solar activity before the Maunder Minimum instead of a sudden drop, further supporting the achievements in Vaquero et al. (2011). Note that Miyahara et al. (2021) estimated that the second solar cycle was only slightly smaller than the first one using cosmogenic radionuclides in contrast with our raw group counting from Scheiner (1630, 1651).

Our results provide valuable hints on the operation of the solar dynamo as the transition between a normal activity regime and the onset of the Maunder Minimum seems more gradual after these corrections; unlike the previous knowledge (see Figure 2 and 8). This gradual transition is also observed during the recovery of solar activity at the end of the Maunder Minimum.

Another important result of our group number recount is that the solar cycle that took place in the 1620s now has the standard 11-yr solar cycle shape with a fast (but not sudden) rise and a slow decay. This is visible in Figure 8 (bottom panel) where we show the annual averages of the raw number of groups. Previously, this solar cycle seemed to be a rather short 8-yr cycle (1624–1632) that would rise to maximum in one year according to V16. This solar minimum has been variously located somewhere in June – December 1620 in Neuhäuser and Neuhäuser (2016) and around 1621 in Carrasco et al. (2019b). After our recount it seems to start ~1620 and finish in ~1632 (with the maximum in ~1625). A similar result is obtained if we replace the observations made by Galileo, Cigoli, Cologna, Colonna, Mögling and Gassendi in V16 by the new group counts obtained by Vokhmyanin and Zolotova (2018a), Carrasco et al. (2020), Vokhmyanin, Arlt & Zolotova (2021), Hayakawa et al. (2021) and Carrasco et al. (2021a) from observations made by those observers (Figure 8, top panel). The years of minima obtained in this work agree with those by Usoskin et al. (2021) from annual $^{14}$C data but are slightly different to those found by Miyahara et al. (2021) also from $^{14}$C data since the latter values set those minima in 1622 and 1633.

Finally, we calculated the active day fraction for the solar minimum in 1632 (the solar cycle before the Maunder Minimum) and compared it with modern minima. According to raw data, the upper limit obtained for the solar minimum of 1632 is comparable to the solar minimum of Solar Cycle 24 and 25. The solar activity level computed for 2008–2009 (minimum of Solar Cycle 24) and 2018–2019 (minimum of Solar Cycle 25) once we discard groups whose observed areas are between 10 and 100 millionths of solar disc from Kislovodsk data is comparable to that in 1632. In addition, the solar activity level calculated for 2008 and 2018 after applying the most conservative constraint is comparable to the lower limit value estimated for 1632. Note that the active day fraction value obtained in this work for the solar minimum in 1632 is slightly lower than that found by Carrasco et al. (2021c) for 1709 (one of the most active years in the Maunder Minimum).

This work demonstrates, as other previous studies have done, how sensitive historical observations are to mistakes in translation and interpretation. It also shows the critical need for improving the databases on which the sunspot number index is based. Only this way, we will be able to reach a more complete understanding of the past, present, and future of solar activity.

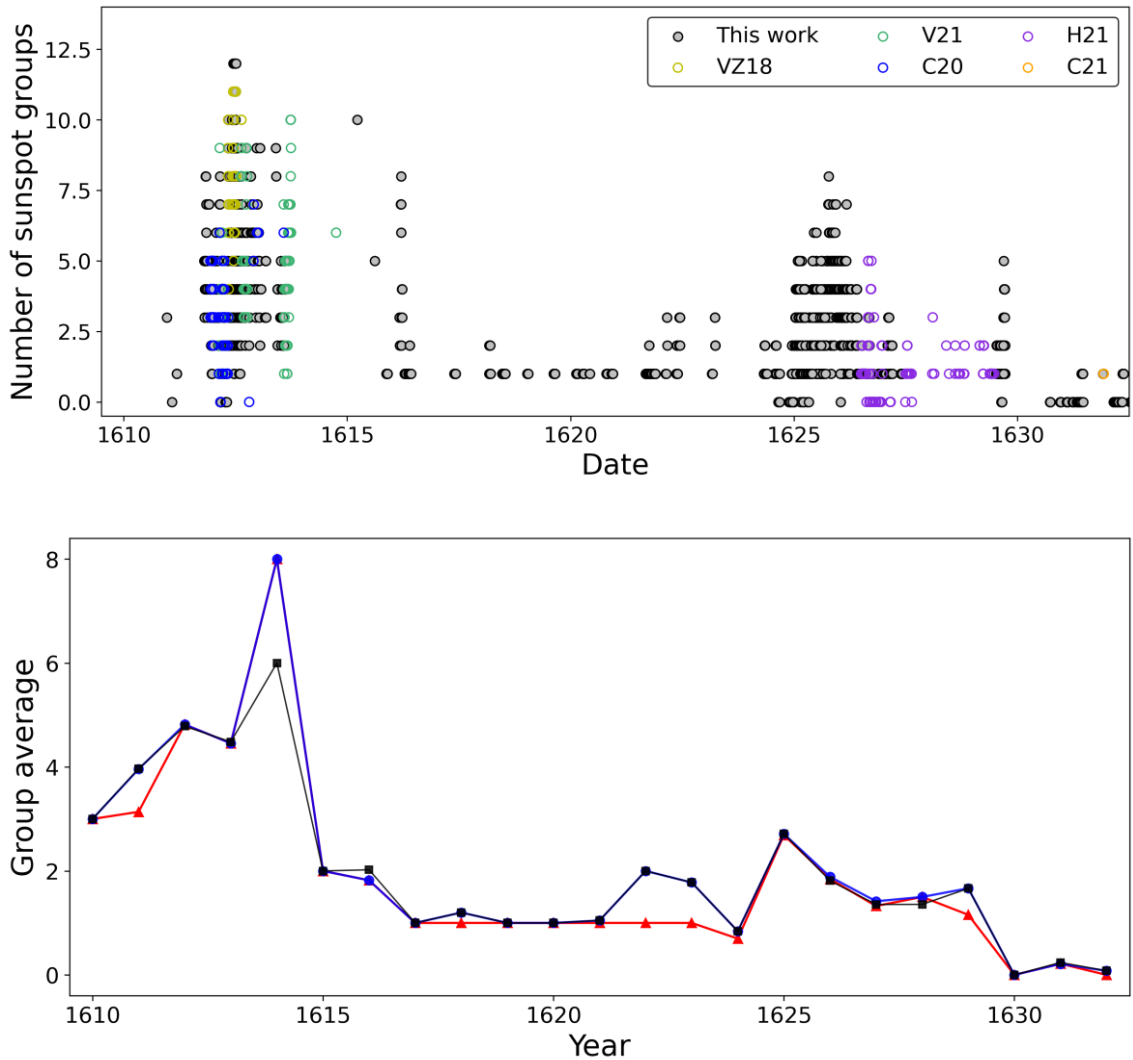

**Figure 8.** (Top panel) Daily number of groups obtained in this work in addition to those including in V16 (see Figure 2, bottom panel) removing data from Galileo, Cigoli, Cologna, Colonna, Mögling, Hortensius, Schickard, and Gassendi (grey dots). New group counts obtained by Vokhmyanin and Zolotova (2018a) (in yellow), Carrasco et al. (2020) (in blue), Vokhmyanin et al. (2021) (in green), Hayakawa et al. (2021) (in violet) and Carrasco et al. (2021a) (in orange) from observations made by these observers are depicted. (Bottom panel) Annual average of the number of groups recorded by the observers included in V16 for the two first solar cycle using telescopic observations (red triangles), that removing Smogulecz's data from V16 and replacing Scheiner's data by the new counting presented in this work (blue circles) and this latter replacing data from Galileo, Cigoli, Cologna, Colonna, Mögling, Hortensius, Schickard, and Gassendi by the new counts obtained by Vokhmyanin and Zolotova (2018a), Carrasco et al. (2020), Vokhmyanin et al. (2021), Hayakawa et al. (2021) and Carrasco et al. (2021a) from observations made by these observers (black squares).

**Acknowledgements**

This research was supported by the Economy and Infrastructure Counselling of the Junta of Extremadura through grant GR18097 (co-financed by the European Regional Development Fund) and by the Ministerio de Economía y Competitividad of the Spanish Government (CGL2017-87917-P). It also has been partially supported by NASA Living With a Star grant NNX16AB77G. The authors have benefited from the participation in the ISSI workshops led by M.J. Owens and F. Clette on the calibration of the sunspot number.

**Disclosure of Potential Conflicts of Interest**

The authors declare that they have no conflicts of interest.

**References**

Arlt R., Senthamizh Pavai V., Schmiel C., Spada F., 2016, Sunspot positions, areas, and group tilt angles for 1611−1631 from observations by Christoph Scheiner, A&A, 595, A104. DOI: 10.1051/0004-6361/201629000.

Arlt, R., & Vaquero, J.M. 2020, Historical Sunspot Records, Living Reviews in Solar 274 Physics, 17, 1. DOI: 10.1007/s41116-020-0023-y.

Carrasco, V.M.S. 2019, Improving sunspot records: misreading of 'Rosa Ursina' by Scheiner, Observatory, 139, 153.

Carrasco, V.M.S., Vaquero, J.M., Gallego, M.C., Villalba Álvarez, J. Hayakawa, H., 2019a, Two debatable cases for the reconstruction of the solar activity around the Maunder Minimum: Malapert and Derham, MNRAS, 485, L53. DOI: 10.1093/mnrasl/slz027.

Carrasco, V.M.S., Gallego, M.C., Villalba Álvarez, J., Vaquero, J.M., 2019b, Sunspot observations by Charles Malapert during the period 1618–1626: a key data set to understand solar activity before the Maunder minimum, MNRAS, 488, 3884. DOI: 10.1093/mnras/stz1867.

Carrasco, V.M.S., Gallego, M.C., Vaquero, J.M., 2020, Number of sunspot groups from the Galileo–Scheiner controversy revisited, MNRAS, 496, 2482. DOI: 10.1093/mnras/staa1633.

Carrasco, V.M.S., Gallego, M.C., Villalba Álvarez, J., Vaquero, J.M., 2021a, A Reanalysis of the Number of Sunspot Groups Recorded by Pierre Gassendi in the Cycle Before the Maunder Minimum, Solar Phys. 296, 59. DOI: 10.1007/s11207-021-01809-1.

Carrasco, V.M.S., Gallego, M.C., Villalba Álvarez, J., Vaquero, J.M., Hayakawa, H., 2021b, Analyses of Early Sunspot Records by Jean Tarde (1615 – 1617) and Jan Smogulecki (1621 – 1625), Solar Phys. 296, 170. DOI: 10.1007/s11207-021-01905-2.

Carrasco, V.M.S., Hayakawa, H., Kuroyanagi, C., Gallego, M.C., Vaquero, J.M., 2021c, Strong evidence of low levels of solar activity during the Maunder Minimum, MNRAS, 504, 5199. DOI: 10.1093/mnras/stab1155.

Chapman, A., 1995, The Astronomical Work of Thomas Harriot (1560-1621), Quarterly Journal of the Royal Astronomical Society, 36, 97.

Clette, F., Svalgaard, L., Vaquero, J.M., Cliver, E.W., 2014, Revisiting the Sunspot Number, SSRs, 186, 35. DOI: 10.1007/s11214-014-0074-2.

Daxecker, F., 2004, The Physicist and Astronomer Christoper Scheiner - Biography Letters, Works (Innsbruck: University of Innsbruck).

Drake, S. 1967, Dialogue Concerning the Two Chief World Systems (Berkeley: University of California Press).

Eddy, J.A., 1976, The Maunder Minimum, Science, 192, 1189. DOI: 10.1126/science.192.4245.1189.

Galilei, G., Scheiner, C., 2010, On sunspots, Galileo Galilei and Christoph Scheiner, 497 University of Chicago Press, Chicago.

Hardy, R., 1991, Theophrastus Observations of Sunspots, JBAA, 101, 261.

Hayakawa, H., Tamazawa, H., Ebihara, Y., Miyahara, H., Kawamura, A.D., Aoyama, T., Isobe, H., 2017, Records of sunspots and aurora candidates in the Chinese official histories of the Yuán and Míng dynasties during 1261–1644, PASJ, 69, 65. DOI: 10.1093/pasj/psx045.

Hayakawa, H., Iju, T., Murata, K., Besser, B.P., 2021, Daniel Mögling's Sunspot Observations in 1626–1629: A Manuscript Reference for the Solar Activity before the Maunder Minimum, ApJ, 909, 194. DOI: 10.3847/1538-4357/abdd34.

Herr, R.B., 1978, Solar Rotation Determined from Thomas Harriot's Sunspot Observations of 1611 to 1613, Science, 202, 1079. DOI: 10.1126/science.202.4372.1079.

Hoyt D.V., Schatten K.H., 1998, Group Sunspot Numbers: A New Solar Activity Reconstruction, Solar Phys., 179, 189. DOI: 10.1023/A:1005007527816.

Keimatsu, M. 1970, A Chronology of Aurorae and Sunspots observed in China, Korea 298 and Japan (Part 1), Annals of science the College of Liberal Arts Kanazawa 299 University, 7, 1.

Kovaltsov, G.A., Usoskin, I.G., Mursula, K., 2004, An Upper Limit on Sunspot Activity During the Maunder Minimum, Solar Phys., 224, 95. DOI: 10.1007/s11207-005-4281-6.

Mandal, S., Krivova, N., Solanki, S.K., Sinha, N., Banerjee, D. 2020, Sunspot area catalog revisited: Daily cross-calibrated areas since 1874, A&A, (in press). DOI: 10.1051/0004-6361/202037547.

McIntosh, P.S.: 1990, The Classification of Sunspot Groups, Sol. Phys. 125, 251. DOI: 10.1007/BF00158405.

Mitchell, W.M. 1916, The history of the discovery of the solar spots, Popular Astronomy, 24, 206.

Miyahara, H., Tokanai, F., Moriya, T., Takeyama, M., Sakurai, H., Horiuchi, K., Hotta, H. 2021, Gradual onset of the Maunder Minimum revealed by high-precision carbon-14 analyses, Scientific Reports, 11, 5482. DOI: 10.1038/s41598-021-84830-5.

Muñoz-Jaramillo, A., Senkpeil, R.R., Windmueller, J.C. et al. 2015, Small-scale and global dynamos, and the area and flux Distributions of active regions, sunspot groups, and sunspots: A multi-database study, ApJ, 800, 48. DOI: 10.1088/0004-637X/800/1/48.

Muñoz-Jaramillo, A., & Vaquero, J.M. 2019, Visualization of the challenges and limitations of the long-term sunspot number record, Nature Astron., 3, 205. DOI: 10.1038/s41550-018-0638-2.

Neuhäuser, R., Neuhäuser, D.L. 2016, Sunspot numbers based on historic records in the 1610s: Early telescopic observations by Simon Marius and others, Astron. Nachr., 337, 581. DOI: 10.1002/asna.201512292.

Otkidychev, P.A., Skorbezh, N.N. 2014, Activity indices in solar cycle 24 and their correlation with general regularities of cycles 19–23 according to mountain astronomical station data, Geomagnetism and Aeronomy, 54, 1014. DOI: 10.1134/S0016793214080131.

Owens, B. 2013, Long-term research: Slow science, Nature 495, 300. DOI: 10.1038/495300a.

Shea, W. R. 1970, Galileo, Scheiner, and the Interpretation of Sunspots. Isis, 61, 498. DOI: 10.1086/350682.

Scheiner, C. 1617, Refractiones Coelestes (Ingolstadt: Typographia Ederiana apud Elisabeth Angermaria).

Scheiner, C. 1621, Oculus, hoc est: Fundamentum opticum (Friburg: Henricum Dulcken).

Scheiner, C. 1630, Rosa Ursina Sive Sol (Bracciano: Andrea Fei).

Scheiner, C. 1651, Prodromus pro sole mobili et terra stabili contra Galilaeum a Galileis (Rome: Societatis Iesu).

Svalgaard, L., Schatten, K.H., 2016, Reconstruction of the Sunspot Group Number: The Backbone Method, Solar Phys., 291, 2653. DOI: 10.1007/s11207-015-0815-8,

Tlatov, A.G., Skorbezh, N.N., Sapeshko, V.I., Tlatova, K.A. 2019, Comparative Analysis of the Catalog of Individual Sunspots Based on Data from Kislovodsk Mountain Astronomical Station, Geomagnetism and Aeronomy, 59, 793. DOI: 10.1134/S0016793219070235.

Usoskin, I.G., Arlt, R., Asvestari, E. et al., 2015, The Maunder minimum (1645–1715) was indeed a Grand minimum: A reassessment of multiple datasets, A&A, 581, A95. DOI: 10.1051/0004-6361/201526652.

Usoskin, I.G., Kovaltsov, G.A., Lockwood, M., Mursula, K., Owens, M., Solanki, S.K. 2016, A New Calibrated Sunspot Group Series Since 1749: Statistics of Active Day Fractions, Solar Phys., 291, 2685. DOI: 10.1007/s11207-015-0838-1.

Usoskin, I.G., Solanki, S.K., Krivova, N., Hofer, B., Kovaltsov, G.A., Wacker, L., Brehm, N., & Kromer, B. 2021, Solar cyclic activity over the last millennium reconstructed from annual $^{14}$C data, A&A, DOI: 10.1051/0004-6361/202140711.

Vaquero, J.M., Gallego, M.C, & García, J.A. 2002, A 250-year cycle in naked-eye 328 observations of sunspots, GeoRL, 29, 58. DOI: 10.1029/2002GL014782.

Vaquero, J.M., Vázquez, M., 2009, The Sun Recorded Through History, Springer, Berlin.

Vaquero, J.M., Gallego, M.C., Usoskin, I.G., Kovaltsov, G.A. 2011, Revisited sunspot data: A new scenario for the onset of the Maunder minimum, ApJL, 731, L24. DOI: 10.1088/2041-8205/731/2 /L24.

Vaquero, J.M., Svalgaard, L., Carrasco, V.M.S., Clette, F., Lefèvre, L., Gallego, M.C., Arlt, R., Aparicio, A.J.P., Richard, J.-G., Howe, R., 2016, A Revised Collection of Sunspot Group Numbers, Solar Phys., 291, 3061. DOI: 10.1007/s11207-016-0982-2.

Vokhmyanin M. V., Zolotova N. V., 2018a, Sunspot Positions and Areas from Observations by Galileo Galilei, Solar Phys., 293, 31. DOI: 10.1007/s11207-018-1245-1.

Vokhmyanin M. V., Zolotova N. V., 2018b, Sunspot Positions and Areas from Observations by Pierre Gassendi, Solar Phys., 293, 150. DOI: 10.1007/s11207-018-1372-8.

Vokhmyanin, M., Arlt, R., & Zolotova, N. 2020, Sunspot Positions and Areas from Observations by Thomas Harriot, Solar Phys, 295, 39. DOI: 10.1007/s11207-020-01604-4.

Vokhmyanin, M., Arlt, R., & Zolotova, N. 2021, Sunspot Positions and Areas from Observations by Cigoli, Galilei, Cologna, Scheiner, and Colonna in 1612 – 1614, Solar Phys, 296, 4. DOI: 10.1007/s11207-020-01752-7.

Willis, D. & Stephenson, F.R. 2001, Solar and auroral evidence for an intense recurrent 338 geomagnetic storm during December in AD 1128, Ann. Geophys., 19, 289. DOI: 339 10.5194/angeo-19-289-2001.

Willis, D., Wilkinson, J., Scott, C.J., Wild, M.N., Stephenson, F.R., Hayakawa, H., Brugge, R., Macdonald, L.T., 2018, Sunspot Observations on 10 and 11 February 1917: A Case Study in Collating Known and Previously Undocumented Records, Space Weather, 16, 1740. DOI: 10.1029/2018SW002012.

Wittmann, A.D., Xu, Z.T. 1987, A catalogue of sunspot observations from 165 BC to 343 AD 1684, Astron. Astrophys. Suppl. Ser., 70, 83. ADS: 1987A&AS...70...83W.

Wolf, R. 1859, Mitth. uber die Sonnenflecken, 8, 201.

Xu, Z., Pankenier, D.W., Jiang, Y. 2000, East Asian Archaeoastronomy. Historical 350 Records of Astronomical Observations of China, Japan, and Korea (Amsterdam: 351 Gordon and Breach Science Publishers).

Yau, K. K. C., & Stephenson, F.R. 1988, A revised catalogue of Far Eastern 353 observations of sunspots (165 BC to AD 1918), QJRAS, 29, 175.